\documentclass[]{piparticle}
\usepackage{graphicx}
\usepackage{amsmath}
\usepackage{CJKutf8}

\usepackage{color}
\newcommand{\changed}[1]{{\color{black}  #1}}

\usepackage[normalem]{ulem}

\begin{document}

\runningauthor{Authora \itshape{et al.}}  

\title{Challenges of `imaging' particulate materials in three dimensions}

\author{Matthias Schr\"oter\cite{inst1,inst2}\thanks{E-mail: matthias.schroeter@ds.mpg.de}
Chen Lyv,\cite{inst3}
Jiayun Huang,\cite{inst3}
Kai Huang,\cite{inst3,inst4}\thanks{E-mail: kh380@duke.edu}}

\pipabstract{
In this perspective article, we discuss the challenges of imaging assemblies of particles in three dimensions. Starting from a brief motivation for the investigation of particulate materials, we provide an overview of experimental approaches developed for imaging particles. We list the challenges and existing solutions associated with X-ray tomography, one of the standard methods to study statics. Subsequently, we discuss challenges and opportunities arising from emerging new techniques such as radar tracking and `smart' tracers for exploring granular dynamics. We close with a tentative view on the outstanding problems and potential solutions in the future.}

\maketitle

\blfootnote{
\begin{theaffiliation}{99}
   \institution{inst1} Max Planck Institute for Dynamics and Self-Organization, 37077 G\"ottingen, Germany.
   \institution{inst2} Hypatia Science Consulting, 37081 G\"ottingen, Germany.
    \institution{inst3} Collective Dynamics Lab, Division of Natural and Applied Sciences, Duke Kunshan University, 215306 Kunshan, Jiangsu, China.
    \institution{inst4} Experimentalphysik V, Universit\"at Bayreuth, 95440 Bayreuth, Germany.
\end{theaffiliation}
}

\section{Introduction}
\label{sec:overview}

\begin{center}
{\it Nothing is built on stone; \\ all is built on sand, \\ but we must build as if the sand were stone.} 
\end{center}

Written in the book \textit{In Praise of Darkness} \cite{Borges1974}, the above poem by Jorge Luis Borges, a legendary writer from Argentina, vividly describes the influence of sand grains on the civilization process happening throughout the past centuries, and continuing till now \cite{Beiser2018}. According to European Aggregates Association, aggregates (e.g., sands, pebbles, and grains) are by far the most used raw materials in the world with a demand of about $6$ tonnes per capita per year \cite{UEPG2021}. Despite the ubiquity, the question of \textit{how to} `build as if the sand were stones' actually touches one major challenge in understanding particulate materials: How to describe them as continuum. 

\begin{figure}
\begin{center}
\includegraphics[width=0.45\textwidth]{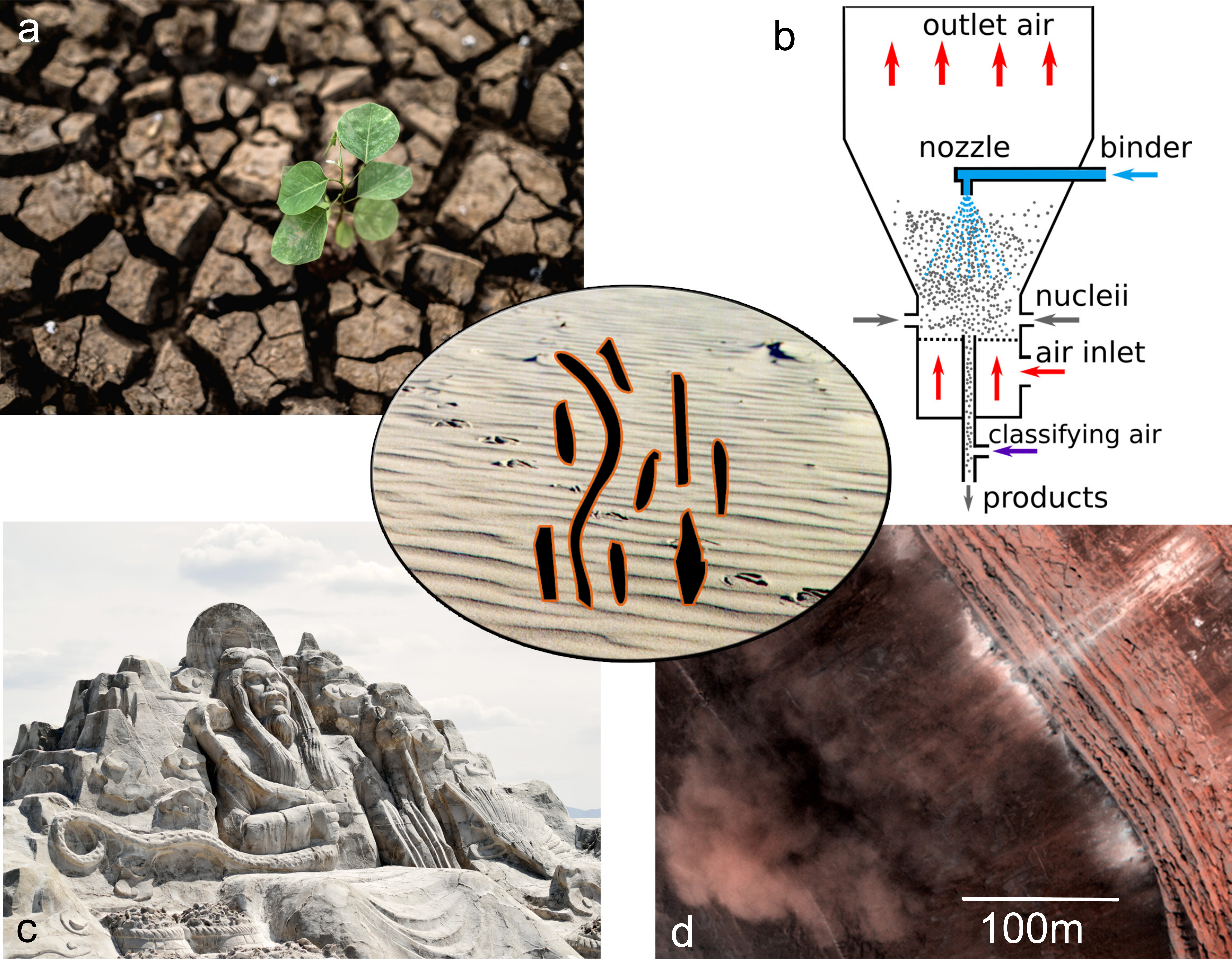}
\end{center}
\caption{A combination of `microscopic' sand grains and `macroscopic' sand ripples forms the character `sand' in Chinese (overlaid in the middle) as appeared in bronze inscriptions over three millennia ago. The ubiquity of particulate materials determines their widespread applications in diverse disciplines, such as (a) ecology (image courtesy of Freepik), (b) granulation process in chemical engineering, (c) sculpturing and construction (sculpture at the Chaka salt lake), and (d) geo-science and space exploration (avalanches on Mars) \cite{NASA2011}.} \label{fig:overview}
\end{figure}

From a historical perspective, particulate materials often act as one essential component in different philosophies and world views across cultures. Taking the pictograph in Chinese as an example, The word `sand' in one of its ancient forms, as overlaid in the center of Fig.~\ref{fig:overview}, provides a clue on how ancient Chinese imagine particulate materials: A composition of small dots that mimic sand grains at a `microscopic' level, together with curved lines that represent the ripples formed by the `macroscopic' collective motion of sand grains. It is one of the first few hundreds words used in the early stage of Chinese language evolution roughly 3400 years ago. Besides its influence on linguistic, the word \begin{CJK*}{UTF8}{gkai} `沙’ \end{CJK*} (pronounced as [$Sh\bar{a}$]) touches another challenge in understanding granular dynamics: How to effectively predict and control the collective behavior of particulate materials from the perspective of `microscopic' particle-particle interactions \cite{Huang2014}.

Challenges in deciphering granular statics and dynamics provide an opportunity for physicists to explore the `root' of diverse applications in different disciplines. From the perspective of sustainability, understanding the interactions between plant roots and the surrounding soil sheds light on ecological models \cite{Meron2015} that can predict the consequence of frequent droughts and flooding arising from global warming (see Fig.\,\ref{fig:overview}(a)). In chemical engineering, the granulation process (see Fig.\,\ref{fig:overview}(b)) is a typical example where energy dissipation at particle level determines the efficiency and quality of the final products \cite{Antonyuk2009, Mueller2016}. In this type of highly dynamical, three-dimensional multi-phase flow systems, it is extremely helpful to follow tracer particles for an adequate calibration of computer simulations, which can in turn make essential predictions to enhance efficiency and to avoid accidents~\cite{Amon2017, Rosato2020, Yule2022}. Making particle-particle interactions attractive, e.g. by adding small amounts of liquids, makes particulate materials perfect for sculpturing, as illustrated by the sculpture made of salt grains (see Fig.\,\ref{fig:overview}(c)). Finally, the frost avalanche observed on Mars (see Fig.\,\ref{fig:overview}(d)) is an another example demonstrating the link between granular solid-liquid-like transition \cite{Andreotti2013} and geo-sciences. Understanding this connection sheds light not only on the triggering mechanism of natural disasters such as earthquakes, snow avalanches, and landslides \cite{Johnson2005, Gray2018}, but also on the evolution of planetary landscape for understanding planet formation and for future space exploration \cite{Jerolmack2019, ruiz-suarez_penetration_2013, Huang2020, Feng2022}. In short, the development of granular physics in the past decades has helped building a framework that can effectively connect diverse research disciplines and relevant industrial sectors. This process has fostered exchange of knowledge and techniques across disciplines from theoretical, numerical and experimental perspectives.

Describing particulate materials in solid, liquid, or gas like states as continuum is a major challenge that has triggered the development of various experimental techniques to `see through sand'. One of the difficulties in formulating such continuum descriptions, is the  energy dissipation at the `microscopic' particle level due to inelastic collisions and friction. Consequently, external energy injection is needed to maintain a fluid-like state: granular fluids are typically driven out of thermodynamic equilibrium and thus statistical mechanics tools cannot be readily used \cite{Jaeger1996, Duran2000, Andreotti2013}. However, there have been substantial progresses in establishing theoretical frameworks for the dynamical and static behavior of particulate materials: From dilute gases of smooth and spherical particles \cite{Haff1983, Haff1983, Poeschel2003, Goldhirsch2003, Eshuis2010, Fullmer2017} to dense granular flow \cite{MiDi2004, Jop2006}; from scaling analysis \cite{Bak1987, Bak1988, Huang2015} via Edwards statistical mechanics \cite{Edwards1989, Baule2018} to quantitative model for non-local rheology \cite{Henann2013, Kamrin2018, Clark2020}. In parallel, there have been substantial efforts in optimizing various numerical approaches to understand granular flow from both fundamental and applications perspectives (see, e.g., a recent editorial  \cite{Aguirre2021}). 
There are excellent reviews and monographs in this direction (see, e.g., \cite{Poeschel2005, Zhao2016, Ge2017, Kieckhefen2020}) along with well established open source simulation packages for simulating particulate materials, such as LAMMPS, MecuryDPM, LIGGGHTS \cite{Thompson1995, Kloss2012, Weinhart2020} for interested readers to explore further.

The development of theoretical and numerical methods is closely associated with lab experiments, or lab simulations as called in some disciplines (see e.g.~\cite{Scheiwiller1987}). The latter plays an indispensable role in testing theories and calibrating numerical models at both `\textit{microscopic}' particle-particle-interaction and `\textit{macroscopic}' collective behavior levels, see, e.g.~\cite{Poeschel2005, Aranson2017}. 

Granular materials are typically opaque and challenging to be 'seen' through. In a previous overview article \cite{Amon2017}, techniques to acquire particle properties in both two- (2D) and three-dimensional (3D) systems were summarized according to their applications. Detailed descriptions of corresponding techniques were provided in the accompanying special issue \cite{Agudo:17,Amon2017_DWS,Born2017,daniels:17,Dijksman2017,Ott2017,Parker2017,Stannarius2017,Weis2017}. Here, those imaging techniques, along with newly developed ones in recent years, are categorized based on exploring static or dynamical properties of particulate materials with in-depth discussions on three representative examples. In light of the overarching goal of this focus series, we focus on challenges and opportunities associated with imaging techniques in the two categories.

\section{A classification of `imaging' techniques}
\label{sec:class}

\begin{figure}
\begin{center}
\includegraphics[width=0.45\textwidth]{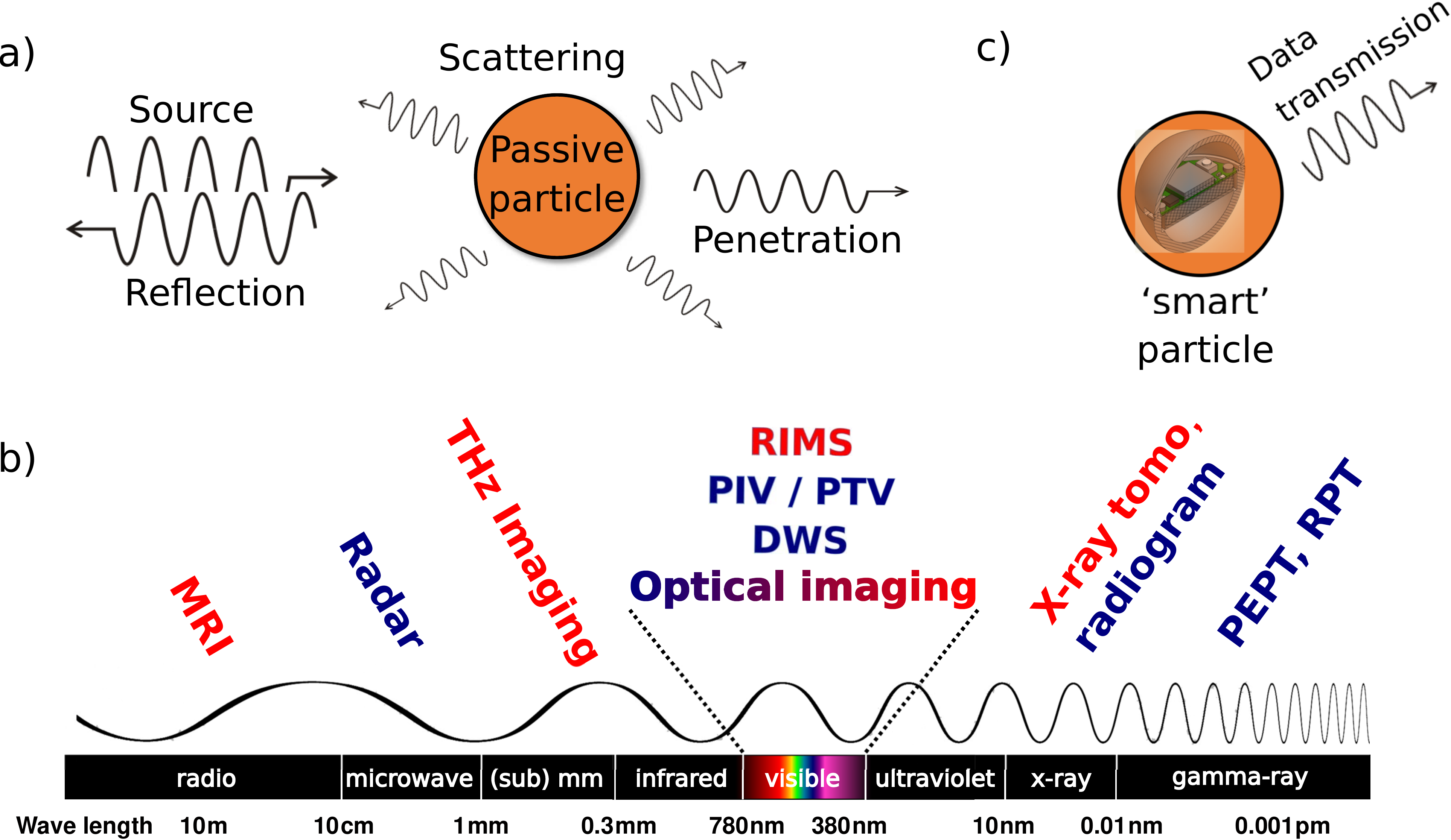}
\end{center}
\caption{(a) A sketch illustrating physical mechanisms associated with `imaging' particles. (b) Experimental methods to probe static (blue text) or dynamic (red text) or both types of properties (blue transitioning to red) of particulate materials in the bulk, by the wavelength of the electromagnetic radiation being used. (c) A `smart' tracer capable of monitoring local properties and transmitting digital data via wireless communications.} \label{fig:tech}
\end{figure}

We first provide an overview of the available techniques to `see' through particulate materials in 3D; in doing so we will use the word imaging in a quite liberal sense. As sketched in Fig.\,\ref{fig:tech}(a), those techniques typically involve sending electromagnetic (EM) waves that effectively penetrate the sample. Based on an analysis of the reflected, scattered, or penetrating signal, one collects local information such as position and velocities of the  particles. Most of the techniques can be sorted according to the frequency of the EM waves being used, as shown in Fig.\,\ref{fig:tech}(b).

In addition to EM waves, acoustic waves have also been used as a non-destructive evaluation (NDE) tool for probing, e.g.~shear-jamming in dense suspensions \cite{Han2016}, precursors of granular failure \cite{brzinski:18} or the network topology of force chains \cite{bassett:12}. 
Last but not least, Fig.\,\ref{fig:tech}(c)  illustrates a new technique involving `smart' tracers with embedded sensors capable of collecting and transmitting digital data through on-board micro-controllers and wireless communications. This approach falls into a different category because EM waves are not used for probing, but rather for data transmission.

For static packings, scanning techniques such as magnetic resonance imaging (MRI) \cite{Stannarius2017}, X-ray tomography (X-Ray Tomo) \cite{Weis2017}, and refractive index matching scanning (RIMS) \cite{Dijksman2017}  can provide the details of the internal structure of a granular packing. However, it is not always straightforward to extract the physical quantities from the large amount of raw data provided by these three-dimensional techniques. Moreover, all of these techniques are comparatively slow, which limits their ability to capture fast dynamics. This topic will be further elaborated in subsection \ref{sec:img}\ref{sec:stat}

Scattering methods have also been used to determine the internal structure of granular media. Given the size range of granular particles (tens of microns to a few millimeters), terahertz waves are the capable of detecting internal structures and their changes in response to external disturbances\cite{Born2017}. By means of diffusion wave spectroscopy with visible lights, we can now detect spatially resolved internal structure changes, thereby paving the way to characterize transient dynamics \cite{Amon2017_DWS}. 

Optical imaging is the most widely used technique to explore granular dynamics in applications ranging from chemical engineering, through geo-science and techniques to space exploration. It is readily available and typically no special treatments of the particles is needed for the analysis. However, due to the opacity of the sample, optical imaging can only characterize the dynamics of particles on the surface or close to the surface in dense systems \cite{Agudo:17} with one exception: If the particles are transparent, and they can be surrounded by a liquid of the same index of refraction, the light sheet scanning technique (RIMS) can be used to observe the bulk of the sample \cite{Dijksman2017}.

To explore granular dynamics in the bulk, tracer particles are typically used. This can be done by means of positron emission particle tracking (PEPT) \cite{Parker2017, Windows-Yule2022}, radio particle tracking (RPT) \cite{Lin1985}, magnetic particle tracking (MPT) \cite{Neuwirth2013, tao:20}, or radar particle tracking \cite{Ott2017, Rech2020}. Challenges associated with particle tracking techniques will be discussed in subsection \ref{sec:img}\ref{sec:dyn}

Different from existing  reviews \cite{Amon2017, Rosato2020}, the goal of this perspective article is to focus on challenges that are not yet fully resolved, based on which potentially promising solutions are discussed to pave the way for further investigations of particulate materials and beyond.

\section{Challenges and Opportunities of Imaging Particles}
\label{sec:img}

\subsection{Spatially resolved 3D imaging}
\label{sec:stat}

\begin{figure*}[t]
\begin{center}
\includegraphics[width=\textwidth]{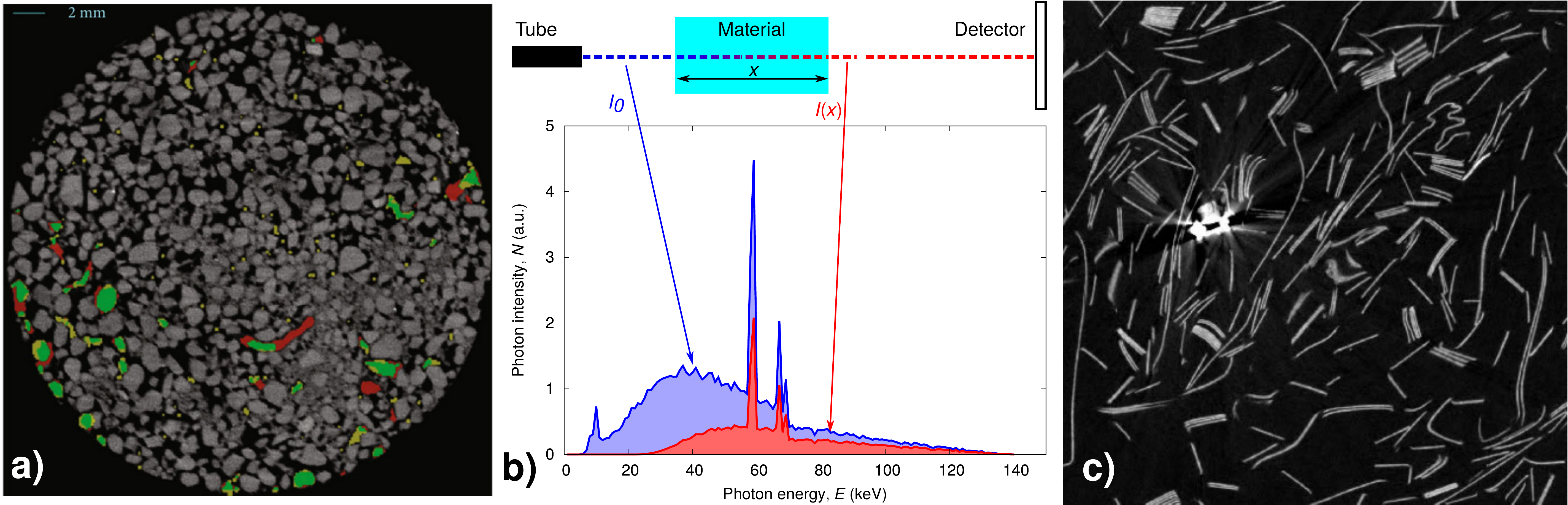}
\end{center}
\caption{Challenges in X-ray imaging.
a) Material mixtures, which do not differ significantly in their attenuation coefficients, can only be segmented by error-prone subsequent image processing. The panel shows as an example a cross-section through a container with plant roots in soil using two different algorithms to identify the root: green areas are roots found by both algorithm, red and yellow areas are found by only one of the two (adapted from \cite{gerth:21}). b) 
Beam hardening arising from the energy dependent attenuation of X-rays: The photon spectrum in front and behind a sample differ qualitatively. This change is not accounted for by the algorithms presently used to compute tomographic images (adapted from \cite{baur:19}). 
c) The panel shows a cross-section of a tomography of paper clippings. The white area with the radially extending white streams is a beam hardening artifact originating from (but not identical to) a metal paperclip contained in the sample. This type of beam hardening artifact occurs if the sample contains particles made from a material with a higher periodic number than the rest.   
} 
\label{fig:statics}
\end{figure*}

The fundamental challenges in acquiring spatially-resolved 3D images from the bulk, i.e.~the inner part, of a dense granular system using electromagnetic waves include: 

(A) The amount of information that has to be gathered from the sample to create  a full  three-dimensional image. If our desired resolution is $m$ micrometer per pixel and our region of interest has a side length of $l$  micrometer, we need to take images with $(l/m)^2$ pixels. Now if we want to measure a cubic volume of  $(l/m)^3$ voxels (the 3D equivalent of pixels), we need $\alpha l/m$ images. The pre-factor $\alpha$ ranges from 1 for RIMS, where we can directly acquire neighboring slices, to $\sqrt{2}$ as frequently used in X-ray tomography. In practice, $l/m$ typically ranges from 500 to 2000. Acquiring this number of individual images is time-consuming. Therefore 3D imaging techniques are well suited to study static granular packings but will have problems to image fast
granular dynamics where particles move with speeds above $\approx$ 0.01 diameters per second. This definition encompasses granular phenomena such as dynamical heterogeneities in slowly driven systems, shear in split bottom cells, dense flows in chutes, silos or down inclined planes, avalanches and segregation in rotating drums, or granular gases driven my electromagnetic shakers.  

This imaging speed problem does not occur if the dynamic process can either be repeatedly stopped and restarted such as e.g.~with grain breakage under load \cite{ando_13,okubadejo_17}. Or if the dynamics happens on the timescales of several hours, i.e.~much longer than the typical acquisition times described below. Examples for this situation include creep under constant load \cite{ando_19}, root growth \cite{viggiani_14,martins_20} or the formation of ice lenses in frozen soil \cite{viggiani_14}.

(B) The interaction between the waves and the material the grains are made from should be material-specific and enable the localization of the measured signal. The latter condition is met by some types of absorption or local excitation. In contrast, refraction will not work for imaging because it will occur at all grain interfaces the waves encounter. In consequence, the grains at the container boundary will blur any image of deeper layers. 

In order to make the subsequent discussion more substantial, we will now focus on X-ray tomography \cite{buzug:08} as imaging modality.  

Regarding A): First, each imaging technique must strike a balance between the signal-to-noise ratio (SNR) of the images and the time allowed to acquire an individual image.  
We can measure the SNR by taking a series of static images and computing the gray value distribution of a pixel in the region of interest. 
The mean of that distribution is the signal we are interested in, and the standard deviation around that mean is the measure for the noise in any individual image. 
Because the generation of photons, their interaction with the sample material, and their probability of being detected in the camera are all stochastic processes, the number of photons detected by any given camera pixel will follow a Poisson distribution.
For Poisson distributions the SNR is proportional to $\sqrt{N}$, where $N$ is the number of photons.

Due to its relevance for our own vision, mankind has developed quite powerful sources of photons in the optical range. This allows optical high speed cameras to acquire hundreds of thousands of images per second at good quality. Photon sources for X-rays have only been explored since the work of Wilhelm R\"ontgen in 1895. As discussed below, most of the available sources are comparatively dim. Combined with the fact that the imaging mechanism is absorption, which further decreases $N$, a reasonable SNR requires often exposure times which make it impossible to image moving objects without motion blur. Even worse, the high dynamic range necessary for absorption imaging will often require the acquisition of $n$ images of the non-moving sample. Averaging those images will then increase the SNR by $\sqrt{n}$, however at the cost of increasing the acquisition time by a factor of $n$. 

In the following we discuss the different types of X-ray tomography setups and the trade-off between x-ray intensity and focal spot size. The former determines the possible acquisition speed, the latter the achievable resolution.  As a rule of thumb, a resolution of 10 or more voxels per particle diameter, is necessary for a reliable detection of the particles.

A frequently used type is commercially built in-house tomography setups. They typically  require exposure times  in the range of 0.1 to 5 seconds for each of the individual radiograms, depending on the material and the desired image quality. For the reconstruction of a 3D tomogram between 1400 and 2800 radiograms have to be taken while the sample is rotated  by 360$^\circ$ in total. Including time for the rotation itself, typical scan duration range from 10 minutes to 4 hours.

As described above, the limiting factor is the X-ray photon flux which comes from a small spot at the anode: the size of this spot has to be smaller than a pixel size in order to avoid blurry images. Given that  98 \% of the kinetic energy of the electrons impinging on the anode is converted to heat instead of X-ray photons, the requirement to not melt the tungsten layer on the anode, which  at the focal spot limits the available photon flux. 

Taking the Nanotom from GE Sensing and Inspection as an example, it has a X-ray tube with 50 W electrical power and an anode made from synthetic diamond for good heat transfer and low absorption. The focal spot sizes can be as small as 1 $\mu$m; in practice voxels with side lengths down to 3 $\mu$m can be realized with reasonable effort. The scan of a granular  sample composed of  200 $\mu$m diameter glass spheres with a resolution of 5 $\mu$m  per voxel and a side length $l$ of 5 mm requires about 135  minutes \cite{hemmerle:16}.  

A presently emerging way to keep the focal spot size small while effectively removing the heat is the use of a jet of liquid metal as anode. Such X-ray tubes are commercially available from the company Excillum. However, these sources produce mostly low-energy X-ray photons which are more suitable for medical than materials science applications. Moreover, due to the fluid nature of the anode, it will be difficult to integrate these x-ray sources in rotating gantry systems, i.e.~systems where source and camera rotate around a stationary sample (see discussion below). 

The next category are medical tomography setups which trade some resolution for speed.
Their X-ray tubes use a fast rotating metal disks as anode which results in the heat being spread out on a ring with the thickness of the focal spot and a diameter of several cm. This allows the use of electrical power in the kW range, the corresponding 
X-ray photon fluxes permit the acquisition of full 3D images in tens of seconds. Which is still to slow for observing most granular dynamics directly. 

In medical tomography setups, the typical voxel size is in the range  between 0.2 mm  \cite{li:21,yuan:21} and 0.6 mm \cite{boerzsoenyi:12,xia:14}. Thus the resolution of these medical setups rules out experiments with grains smaller than 2 mm in diameter.

One big advantage of medical tomography setups is that they  are rotating gantry systems, i.e.~the X-ray source and camera move around the stationary sample (normally a human patient). This absence of centrifugal forces is a necessary condition for tomographic studies of fast granular dynamics.

The third category of tomography setups is situated at synchrotron sources which provide significantly higher X-ray photon fluxes in a parallel beam geometry. The latter allows to acquire the radiograms  necessary for reconstruction while rotating the sample only by 180$^\circ$, unlike the 360$^\circ$ required in the cone-beam geometry of classical X-ray tubes.  Acquisition times for full 3D images range from 15 s (beamline 2BM at the Advanced Photon Source  at the Argonne National Lab \cite{wang:14}) to 0.2 s (beamline ID15A at the ESRF in Grenoble).
However, taking 5 full tomographies per second, requires to rotate the sample 2.5 times per second. Assuming a container radius of 1 cm, this results in centrifugal forces of 2.5 m/s$^2$ at the sample boundary. While it is possible to suppress the effect of this centrifugal force by applying a strong enough confining pressure on the sample,this confining pressure will also suppress the types of granular dynamics described above.

Finally, there are specialized setups with non-rotating samples and significantly faster acquisition times of 1000 \cite{waktola:18,stannarius:19} to 2500 \cite{homan:15} radiograms per second. But these are limited to only one or two cross sections through the sample and do not provide a full 3D image.  
So in summary, standard X-ray tomography as a method is more suitable to study static samples than granular dynamics.

A speed-up of tomographic imaging can also be achieved if prior information is used for the reconstruction. E.g.~if the initial particle configuration is known from a full 3D scan, a comparison of experimentally measured, time resolved radiograms and synthetic radiograms generated from simulations can be used to describe the evolution of the sample \cite{khalili:17,gupta:21}. Or if the particles in the sample are highly monodisperse spheres of known radius, their apparent size in each radiogram can be used to determine their full 3D positions \cite{ando:21}.

Regarding B, one of the main advantages of X-rays is that their index of refraction deviates only by $10^{-6}$ or less from unity for all interesting solids and fluids. In consequence, X-rays are (almost) not refracted at any particle surface boundary\footnote{For coherent X-ray sources such as synchrotrons and with larger sample to detector distances the remaining small amount of refraction can actually be used to improve the contrast in the images. This method is called phase contrast tomography.}, thus all the particles are so to speak index of refraction matched.
The signal retrieved from the absorption within the sample is integrated along the straight path the X-ray photons travel. There are two important practical considerations regarding the use of absorption as imaging mechanism: beam hardening and sample contrast.

Beam hardening describes the deviation from the text book situation where the X-ray beam is mono-energetic and its  attenuation can be  described by the Beer-Lambert law 
\begin{equation}
 I = I_0 e^{-\mu \Delta x }  
 \label{eq:lambert-beer}
\end{equation}
here $I_0$ and $I$ are the intensities before and after the sample of thickness $\Delta x$ and $\mu$ is the absorption coefficient which depends on the material of the sample. This exponential dependence is the basis for the standard models for X-ray imaging, both of the individual projection images, and the 3D reconstruction derived from those images.  

However, X-ray sources are typically not mono-energetic, they provide photons with a broad energy spectrum $N(E)$, as illustrated by the blue flux curve in figure 
\ref{fig:statics} b). At the same time, the attenuation coefficient $\mu(Z,E)$ is not only a function of the order number $Z$ of the sample atoms, but also of the photon energy $E$. In consequence, the X-ray photon distribution is not only diminished but also quantitatively changed after passing through the sample, cf.~the red curve in figure \ref{fig:statics} b).  Because the weight of higher energy photons (i.e., `harder' X-rays) increases, this effect is called beam hardening.

Taking into account the \changed{energy-dependent} sensitivity  $S(E)$ of the detector, the measured intensity $I$ is described by:
\begin{equation}
I(x) \propto \int N(E) \, \,  \exp\{-\mu(E,Z) \, \Delta x\}  \, \,  S(E) \, \, \mathrm{d}E,
\label{Equ:Int-Integral}
\end{equation} 
which is quite different from equation \ref{eq:lambert-beer}.

In practice, it is challenging to implement \ref{Equ:Int-Integral}. This is not due to the energy dependence of $\mu$, which has been pre-computed by the National Institute of Standards and Technology (NIST) and can be downloaded from their website \cite{berger_nist_2010}. But $S(E)$ and $N(E)$ are typically not known for a given tomography setup.

If the sample is made from one material only, it is possible to go back to equation \ref{eq:lambert-beer} by using an effective attenuation coefficient $\mu_{\rm eff}(E_{\rm max},\Delta x)$ which depends on both the maximal X-ray energy $E_{\rm max}$ and the thickness $\Delta x$ of the material. However, the calibration procedure necessary to determine $\mu_{\rm eff}$ is time consuming and specific to a given experimental setup and sample material \cite{baur:19}. Additionally, $\Delta x$ now occurs explicitly and implicitly (via  $\mu_{\rm eff}$) in the exponent of the Beer-Lambert equation. Determining $\Delta x$ will therefore require iterative methods. This path is only followed for high precision experiments \cite{kollmer:20}.

The main problem in the context of this discussion is that all algorithm, which are used to compute the 3D tomography from the stack of X-ray images taken while rotating the sample, are based on equation \ref{eq:lambert-beer}.
The discrepancy between Eq. \ref{eq:lambert-beer} and \ref{Equ:Int-Integral}  leads to various artifacts in the 3D reconstruction. 
On example is shown in figure \ref{fig:statics} c): if the sample contains particles which are made from elements with higher order number $Z$, the reconstruction will misrepresent their size and contain brighter beams radiating from those particles. Another beam hardening artifact is cupping: objects close to the boundary of the sample container will appear brighter than identical objects in the center of the container. 

The ideal solution for beam hardening is to use mono-energetic X-rays. However, this is only feasible in synchrotrons where due to the abundance of photons it is possible to use crystals as monochromators and still have a sufficient photon flux \cite{xia:15}. For in-house tomography setups, it is good practice to position a filter, such as a sheet of copper with a thickness of a few hundreds of $\mu$m, in front of the anode. This will remove most of the lowest energy photons (i.e., acting as a high pass filter in the spectrum shown in figure \ref{fig:statics} (b)). Consequently, this approach decreases the effect of beam hardening in the sample, sometimes almost completely \cite{ando:21}.

The second point when using absorption as the imaging mechanism is the conceptual question: do we have enough contrast to observe the inner structure of the sample?  The attenuation of X-rays depends on both the number density $\rho$ and order number $z$ of the atoms in the beam path. Consequentially, the answer depends on details of the sample.

If we are interested in a sample of particles with air in the interstitial voids, the difference in $\rho$ will typically provide enough contrast to segment the tomography into a particle and a void phase. If we want to identify the individual grains form the connected particle phase, we need some subsequent image processing. If the particles under consideration are spheres, this separation step is simple because the contact points between individual particles are only a small subset of the overall surface of the particles and therefore easy to identify and remove \cite{Weis2017}. 

If the particles are polyhedra with the possibility of having extended face-to-face contacts, it becomes harder to separate individual particles \cite{neudecker:13,xu:21}. Last but not least, for the case of irregular, real world granular media such as dry sand, more complex segmentation strategies are needed   \cite{vlahinic:14,wiebicke:17}.

If the sample is composed of two or more different materials, the ability to distinguish them depends a lot on their chemical composition, respectively their average $Z$. It can become particularly challenging if different material components share similar attenuation coefficient, such as the plant roots in soil shown in figure \ref{fig:statics} a). Segmentation of these requires either human  interaction \cite{bull:20} or  highly customized algorithms incorporating domain knowledge \cite{gerth:21,anselmucci_21}.

If one of the two materials is a liquid, doping can be a helpful strategy. Studying liquid bridges between plastic spheres is difficult due to the small contrast. However, by using bromodecane, or water with a high concentration of CsCl or KJ dissolved in it,  we can generate 3D images where all three phases: plastic spheres, liquid bridges, and interstitial air, correspond to well separated gray value ranges \cite{weis:19}.  Responsible for this effect are here the high $Z$ components bromine, cesium, or iodine. A similar idea can be used to study the orientation of spherical particles: Covering some part of the surface with a gold or silver layer helps to measure the orientation of individual particles. Such a layer can either be deposited chemically, or by sputtering \cite{hiller:19}.

In addition to identifying particle positions and orientations, resolving the interparticle forces in 3D would be extremely helpful in diverse applications. Doing this with conventional X-ray tomography is difficult because the deformation of most particles is typically too small to get reliable estimates for the forces creating them. RIMS studies of soft hydrogel spheres seem a  more promising venue \cite{Brodu2015}. More recently, the technique combining tomography and measurements the strain tensor of single crystal spheres using diffraction also emerges as an important alternative \cite{hurley:16,Hurley2022}.

\subsection{Time-resolved particle tracking for fast dynamics}
\label{sec:dyn}

Particle tracking is a powerful tool to explore granular dynamics, owing to the capability of continuous tracking in 3D at relatively high temporal resolutions. To ensure sufficient signal-to-noise ratio (SNR), signals emitted or scattered from the tracer particles need to be sufficiently strong to enable tracking. To enhance temporal resolution, continuous operation of the system are typically preferred in comparison to scanning techniques. In practice, multiple detectors are needed to locate tracer particles in 3D. 

\begin{figure*}
\begin{center}
\includegraphics[width=0.85\textwidth]{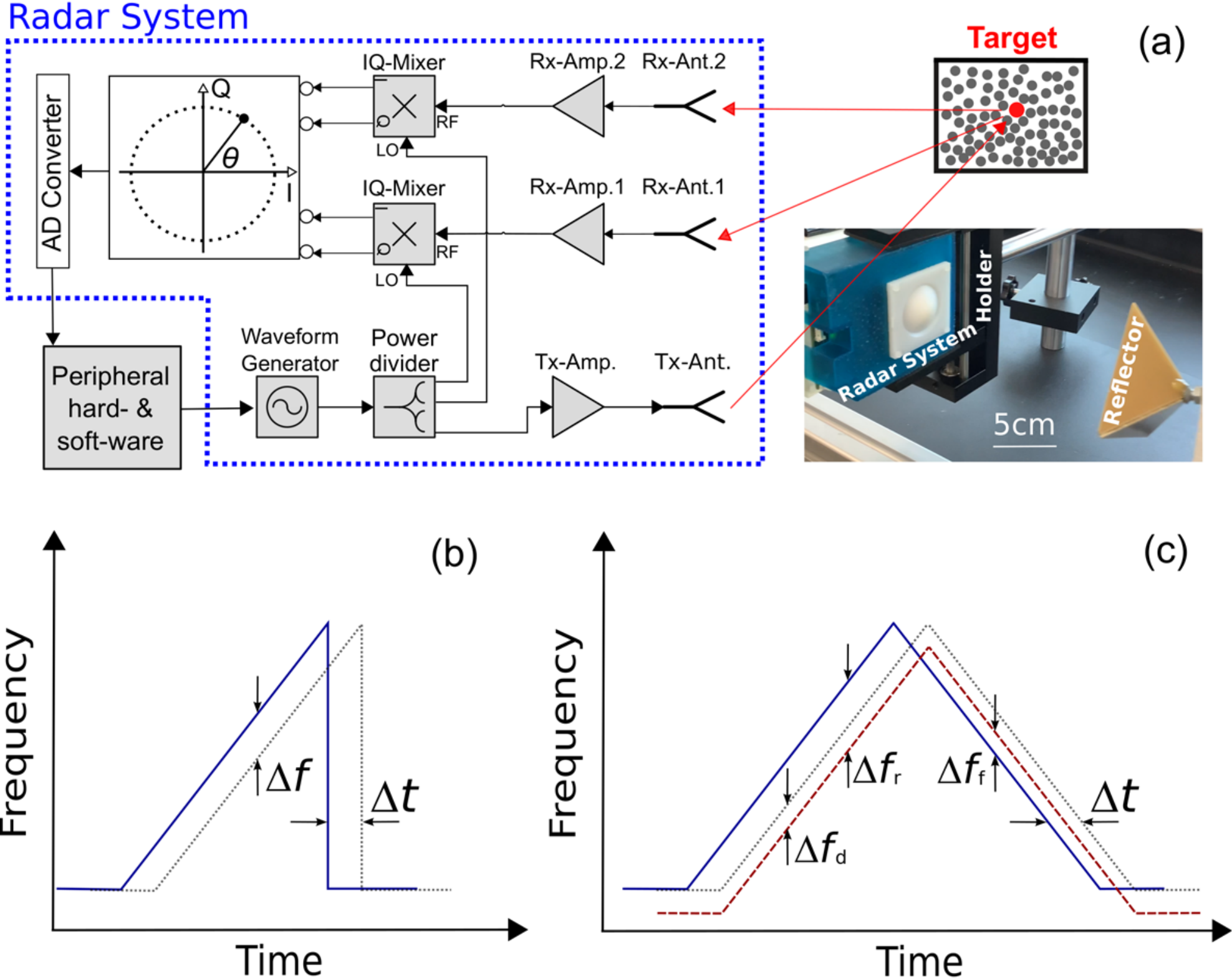}
\end{center}
\caption{(a) A schematic diagram showing the working principle of a multi-static (i.e., transmission and receiving antennae are located differently) radar system. It is adapted from a previous work~\cite{Ott2017} to better present the on-chip radar system (see the picture on the lower right part). The corner reflector shown in the picture is used for calibration. (b) and (c) represent two scenarios for radar tracking in FMCW mode. Blue curves corresponding to the sweeping signal emitted at the transmission antennae, the dashed curves correspond to the received signal from static tracers, and the red curve in (c) represents the received signal in the presence of doppler shift induced by moving targets.
} \label{fig:fmcw}
\end{figure*}

Taking radar particle tracking \cite{Ott2017} as an example (see Fig.\,\ref{fig:fmcw}(a)), it uses the phase shift of EM waves travelling from transmission antennae (Tx-Ant) via tracers to various receiving antennae (Rx-Ant) to identify the relative distance changes of individual targets. Based on the phase shifts, 3D trajectories of the tracers can be constructed through triangulation. The phase shift is extracted using a device called IQ-Mixer, the output of which, plotting in the $I-Q$ plane, gives rise to the phase angle $\theta$. For more information on radar systems in general or radar particle tracking, interested readers may refer to the existing literature \cite{Ott2017,Rech2020, Skolnik2008}.

FMCW (frequency modulated continuous wave) radar is another typically used radar tracking mode. It has the advantage of obtaining both position and velocity of multiple objects through sweeping frequency within the pre-set bandwidth. Two linear frequency chirps, one with a sawtooth form (Fig.\,\ref{fig:fmcw}(b)) and the other one with a triangular form (Fig.\,\ref{fig:fmcw}(c)), are used as emitted signals. Once a chirp signal is chosen, the frequency shift $\Delta f$ between transmission (blue curve) and receiving (dashed gray curve) signals maps directly to the time delay $\Delta t$ in between, and consequently the distance between the radar chip and the target. If the target moves, doppler effect leads to detectable frequency shift $\Delta f_{\rm d}$ added in addition to $\Delta f$. Therefore, a separation of the two contributions leads to both range and velocity information. 

More specifically, as shown in Fig.\,\ref{fig:fmcw}(c), a triangular shaped chirp is used. Assuming that the velocity of the tracer does not change dramatically during the sweeping period, we consider constant $\Delta f_{\rm d}$ induced by the doppler effect. The sketch tells us that range information can be obtained from $\Delta f = (\Delta f_{\rm r}+\Delta f_{\rm f})/2$ and the velocity information can be extracted from $\Delta f_{\rm d} = |\Delta f_{\rm r}-\Delta f_{\rm f}|/2$. Note that Fig.\,\ref{fig:fmcw}(b) and (c) only illustrate the algorithm behind FMCW. In reality, how to accurately determine the position and velocity of multiple tracers requires further signal processing, particularly for the case of tracking multiple targets moving at varying speed. With the latest development of on-chip radar technologies, radar systems are becoming more readily available and compact, as the picture in Fig.\,\ref{fig:fmcw}(a) shows \cite{Saponara2019}.

Challenges associated with this technique include: 

\begin{enumerate}

\item How to select tracer and surrounding particles for a better SNR? Similar to other imaging techniques, the tracer particles should have sufficient contract with respect to the surrounding particles. For the case of EM waves in radar tracking, the dielectric constant of the tracer materials has to be carefully selected for the tracers to be distinguished from their neighbors. Note that sometimes coating a thin layer of conductive materials can enhance the contrast to a satisfactory level.  

\item How to ensure that tracer particles are representative? Tracers typically have different properties than surrounding particles. This difference is likely to generate segregation \cite{schroeter:06,Umbanhowar2019}. To ensure mixing it is important to try to match the density and size of tracers in the host material. 
To tackle this challenge, the aforementioned coating techniques can also be used, if the coated layer is sufficiently thin to avoid detectable change of particle size and density. However, we note that coating is not always a feasible solution, thus how the trajectories of tracer particles represents granular dynamics in the bulk is still a topic for further investigations. 

\item How to properly calibrate the system for accurate tracing? Given a certain initial condition, such as sensor locations and sensitivities, the calibration process yields a set of system dependent parameters for reconstructing the 3D trajectories. These parameters play a key role in determining the accuracy and reproducibility of particle tracking, thus a well defined protocol is needed for the calibration process.

\item How to properly choose the working distance and the field of view? The distance change between tracers and the detectors leads to fluctuations of the received signal amplitudes, which in turn lead to IQ mismatch and consequent error in detecting the phase shift between transmission and receiving signals \cite{Rech2020}. Note that radar systems often use phase signals to obtain relative distance change. The phase signals are represented by the I (imaginary) and Q (real) signals obtained by an IQ mixer (see Fig.\,\ref{fig:fmcw} for a sketch of a radar system). Because any distortion of the IQ signals will introduce uncertainty to the phase signals detected, an appropriate correction of IQ mismatch is essential for radar particle tracking. 

\item How to track multiple particles spontaneously at a reasonable sampling rate? Radar can work in various tracking modes through tuning the form of emitted EM waves. Continuous wave radar \cite{Ott2017, Rech2020} has the advantage of tracking particles with high temporal resolution, while FMCW radar is capable of tracking multiple tracers at the same time. In the latter case, it is necessary to choose sweeping time such that it balances temporal and spatial resolutions. On the one hand, faster sweeps enhances sampling frequency. On the other hand, steeper slope arising from faster sweeps leads to higher uncertainty in determining the distance, thus one always needs to find the balance between temporal resolution and tracking capacity (i.e., number of particles to be traced at the same time).  

\end{enumerate}

\begin{figure}
\begin{center}
\includegraphics[width=0.45\textwidth]{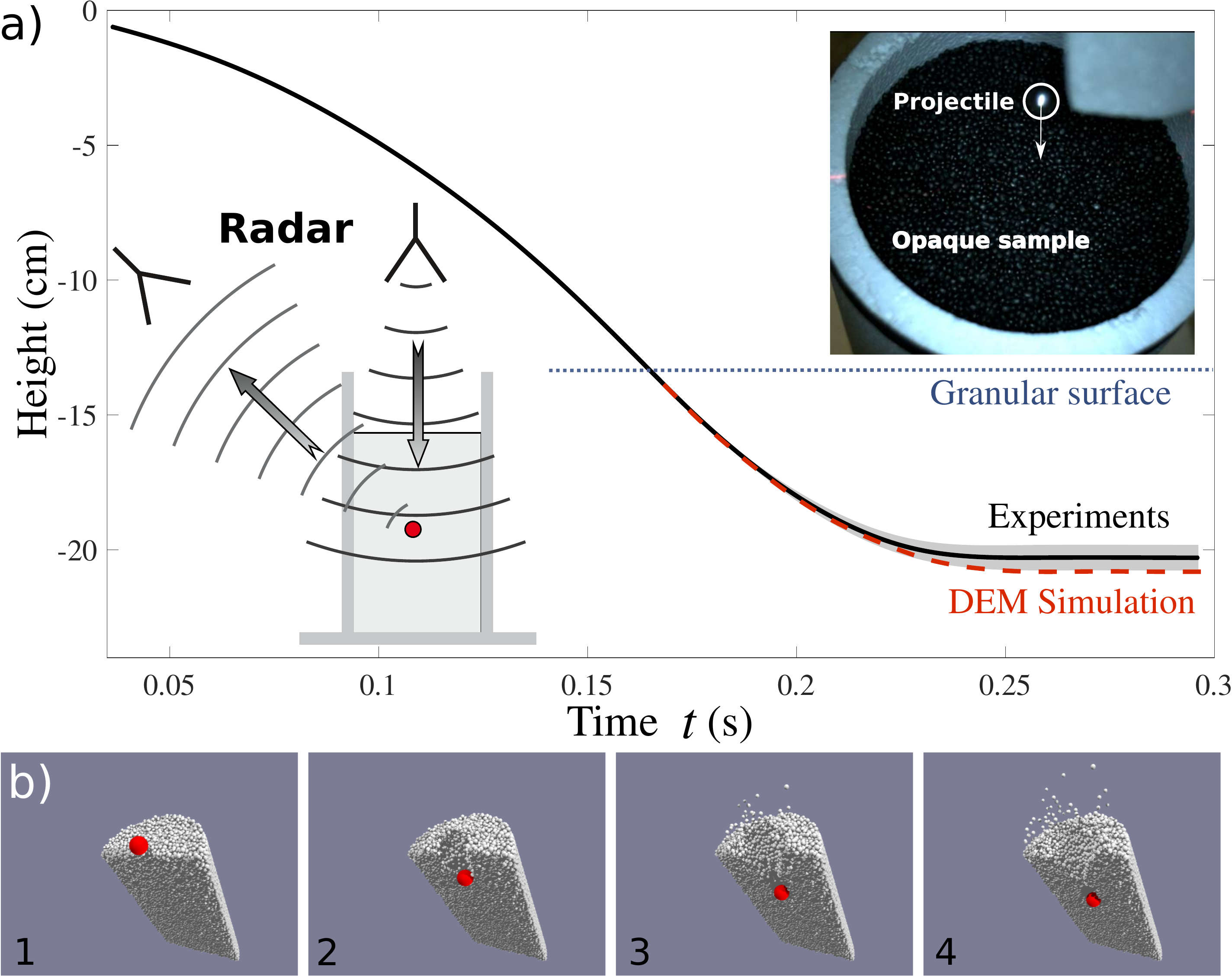}
\end{center}
\caption{(a) The trajectory of a spherical metallic tracer impacting onto opaque Styrofoam particles (see upper right inset for a snapshot of the sample) obtained with a radar setup (see lower left inset for a schematic of the setup) together with the result from numerical simulations. Gray shaded region corresponds to the uncertainty of the trajectory based on multiple runs of experiments. (b) Projectile (red) impact on a granular material composed of spherical particles (silver). Data extracted from the DEM simulation of a corresponding experiment \cite{Huang2020}. } 
\label{dynamics}
\end{figure}

Figure~\ref{dynamics} shows one application using radar particle tracking. A metallic sphere is used as the tracer to measure the drag force exerted by the surrounding particles. Due to the opacity of the sample, it is necessary to have a non-invasive 3D tracking approach to explore what happens within the granular sample. A continuous wave radar system \cite{Huang2020} was used to fulfill the task. Although moving target indicator (MTI) radar is typically used in scenarios at large length scales (e.g., assist aircraft landing or taking-off), it is possible downsize it to the laboratory scale to trace centimeter or even smaller sized particles \cite{Ott2017}. In this specific case, the above challenges can be resolved by means of (i) selecting tracer and surrounding particles such that they have very different dielectric constant to ensure good SNR. More specifically, porous particles can be used to ensure good penetration of EM waves. (ii) Calibration is performed with tracers moving in a well defined circular trajectory. Typically, multiple runs of the calibration process are needed to ensure an accurate determination of calibration parameters such as the positions and orientations of individual antennae. In addition, polarization of the antennae should also be adjusted during the calibration process. (iii) Depending on specific applications, we may choose continuous wave (CW) radar for extremely fast dynamics, while switching to FMCW mode for tracking multiple particles. 

In both SNR enhancement and IQ mismatch correction, how to effectively and robustly smooth raw data collected at different sampling rate and uncertainty levels requires substantial care. Given the fact that the trajectories of tracers are unknown, fitting the raw data with pre-defined trajectories and parameter settings won't help in extracting relevant information. Thus, non-parametric algorithms such as moving average filtering, smoothing splines or local regression are feasible approaches. In particular, Gaussian Process Regression (GPR) has emerged as a suitable method with the ability to handle data with complicated patterns. As a machine learning (ML) algorithm, it trains a model with the collected data for prediction  \cite{Rasmussen2006}. The advantage is that it does not rely on a pre-defined analytic expression, but on the Gaussian distribution of ordinate values at different abscissa data points. Nevertheless, one needs to carefully evaluate computational cost spent in training the model while implementing GPR. Previous investigations \cite{Rasmussen2006} show that GPR has a space complexity of $N_i^3$ and time complexity of $N_i^2$ with $N_i$ the number of data points. For the example described above, $N$ may easily reach $10^5$, therefore an appropriate choice of the kernel function and sampling rate for data smoothing to balance efficiency and accuracy is necessary.

For the projectile impacting example described above, an accurate determination of the tracer trajectory does not necessarily mean that we are ready to explore the collective dynamics of surrounding particles. In this regard, synergy with numerical simulations becomes essential. As shown in Fig.\,\ref{dynamics}(a), direct simulations of the experimental condition provide an opportunity to explore the dynamics of all relevant particles during impact. On the one hand, experimental data can be used to calibrate DEM simulations (see, e.g., Fig.\,\ref{dynamics}(b)). On the other hand, DEM simulations provide an indispensable insight of the experimental results with great details. A calibrated numerical model enables further analysis of mechanical response of the system from a `microscopic' perspective.

Although radar tracking is relatively new in the family of `imaging' particles, it has the potential to develop itself as a compact, cost-effective, energy efficient, and accurate particle tracking approach because of recent advances in radar-on-chip (RoC) technologies. Nowadays, RoC has become widely used in applications such as autonomous driving and gesture recognition \cite{Saponara2019}. Modern on-chip radar effectively embeds transmission antennae (Tx-Ant.), receiving antennae (Rx-Ant), oscillators, amplifiers and other components into a single micro-chip with surface area smaller than $7 \times 7$mm$^2$. Together with micro-controller and other peripheral components, it it possible to embed the whole system into a hand-hold device, as Fig.\,\ref{fig:fmcw}(a) shows. Depending on the bandwidth it operates, commercial models of on-chip radar are capable of measuring centimeter sized tracer at a spatial resolution of centimeter scale within a range of $10\sim 20$ meters. The working distance arises from the damping of EM waves in the sample as well as from the capability for the target to scatter EM waves (characterized by radar cross-section), as indicated by the Radar equation \cite{Skolnik2008}. As such, the minimal size of the target and working range of the radar depend on the power, operation wavelength, cross-section of the tracer, as well as the surrounding environment. For on-chip radar with limited power consumption, further considerations are needed to implement it in particle tracking. The wavelength of commercially available on-chip radar system is on the order of millimeters, thus it theoretically possible to track particles of millimeter size or even smaller. Note that in a previous investigation \cite{Ott2017}, tracking a tracer of $5$\,mm in diameter using EM wave of wavelength $3$\,cm was achieved.

\subsection{`Smart' particles}
\label{sec:smart}

Apart from the methodologies described above, `smart' particles with embedded sensors (see Fig.\,\ref{fig:tech}(c)) have arisen as an alternative for probing the interior of granular sample. Along with the development of MEMS (micro-electromechanical systems) technologies, sensors such as IMU (inertial measurement units) capable of measuring angular velocity, acceleration, magnetic fields, and other properties can be encapsulated in a microchip with a typical size of a few millimeters for convenient integration into the peripheral circuit with a micro-controller. Since the pioneer use of accelerometers in tracing projectile motion inside a granular medium \cite{Altshuler2014}, there have been substantial progresses in developing more robust and accurate tracking algorithms for the detection of tracer position and velocity in both translation and rotational degrees of freedoms by means of sensor fusion (see, e.g., \cite{Koestler2021}). As IMU sensors become more and more widely used in consumer products, they become more compact, easier for system integration and data transmission, and more readily available. This progress sheds light on further development of `smart' tracers for comprehensive tasks with broader applications in industries (see e.g., \cite{Swarm2019}). This technique is particularly useful in conditions where conventional 3D imaging techniques cannot be easily implemented, e.g., in microgravity experiments or in exploring inaccessible regions where space, power, and safety limitations arise. In those scenarios, the possibility of having `smart' particles capable of `tracing' themselves and their surroundings, as well as sending data via wireless communications extends our ability to `see' through particulate materials. 

In practice, there are three main challenges in accurately determining the position of the `smart' tracers: 

\begin{enumerate}
    \item Coordinate transformation: Raw data collected from IMU sensors include angular velocities and accelerations measured in sensor coordinate. Step-by-step coordinate transformation into a fixed laboratory frame is necessary as the moving target may rotate while traveling. For a successful reconstruction, accurate synchronization of accelerometer and gyroscope during the sampling process, removing drifting of the sensor output, as well as an appropriate selection of sampling rate need to be carefully considered \cite{Koestler2021}. 
    
    \item Cumulative error: After obtaining acceleration in a fixed coordination system, integration is needed to obtain velocity and position of the tracer. During this process, error arising from background noise, drift of the raw signal, as well as uncertainty from coordinate transformation, will accumulate. It is recommended to use techniques such as a combination of forward and backward integration to reduce accumulated error \cite{Koestler2021}. 
    
    \item Calibration: While talking about particle tracking, we typically consider measuring geometrical center or center of mass. In either case, the goal is to obtain the trajectory of a fixed point of the tracer particle. Nevertheless, the embedded sensors are not necessarily at the desirable location and different sensors might be located at different locations. Consequently, the acceleration measured might include, e.g., centrifugal acceleration components. Thus, calibration for sensor position is needed for an accurate reconstruction of the tracer trajectory.      
\end{enumerate}

\noindent For more specific challenges associated with individual steps described above, interested readers may refer to a recent article \cite{Koestler2021}. In the future, the development of sensing technology allows more information, such as local humidity and temperature, force applied on the surface, magnetic field, to be collected as `smart' particles going through the sample. It is possible to design `smart' rotors that can effectively explore the sample by themselves actively, in addition to the swarmed tracers described above \cite{Swarm2019}. Moreover, `smart' tracer network with communications and data sharing among each other provides the opportunity for the implementation of data analytical approaches to enhance the `smartness' of the sensing network. In short, if power consumption of the sensing units is further reduced, this technology has a great potential to develop itself into a powerful tool to explore particulate materials from an `insider' perspective.

\section{Outlook: Potential Solutions of Standing Challenges}

To summarize, there have been substantial progresses in the development of numerical and experimental techniques to decipher particulate materials from both local and global perspectives. Throughout the process, a frequently raised question is how the two approaches can better support each other. On the one hand, as described in subsection~\ref{sec:img}\ref{sec:dyn}, numerical simulations are capable of providing all the details of a sample. This is an important extension to current experimental approaches to `see' through particulate materials. On the other hand, we need to ensure the model and relevant parameters are properly chosen to match experimental conditions to empower the conclusions drawn from numerical simulations in widespread applications. As such, techniques for `seeing' through particulate materials become essential in developing models for numerical simulations and in determining relevant parameters for specific scenarios, such as pattern formation ~\cite{bizon:98, boerzsoenyi:09, butzhammer2015}, gravity driven flow, mixing and segregation~\cite{schroeter:06,tsukahara:08, boerzsoenyi:09}, and other applications~\cite{bannerman:11, boerzsoenyi:12, Huang2018, Wu2018}. 

In the future, we anticipate more synergy between 3D imaging techniques and computer simulations at the level of individual particles. In addition to the radar particle tracking example described above~\cite{Huang2020}, recent investigations also show that micro-CT data can be used to calibrate DEM simulations~\cite{moreno-atanasio:10, Dosta2018}. Along with the emerging new techniques to `visualize' local contact forces, we will be better positioned to develop numerical tools for better prediction of the collective behavior of particulate materials.

We also see great potential in applying machine learning (ML) algorithms, especially deep neural networks, in the acquisition of 3D images. The reasons for this optimism are twofold: (i) The large data sets associated with 3D imaging become a boon as they can be used  to train better, more complex models. (ii) While granular dynamics is hard to describe at a continuum level, we know that there are underlying generative models such as the Newtonian dynamics of the individual grains or the condition of non-overlap between particles. While it is hard to write those down explicitly for even a medium sized sample of grains,
neural networks excel in finding and utilizing such generative models (in an implicit manner though).
See, for instance, recent examples of the successful application of ML to 3D image acquisition are particle detection algorithm in granular gases using imaging \cite{Puzyrev2020}  and positron emission particle tracking \cite{nicusan:20}. 

Further on, some ML algorithm can directly improve the raw images taken by e.g. decreasing the noise level \cite{lehtinen:18} or correcting  beam hardening artifacts in X-ray images \cite{cong:21}. Moreover, the presently arriving generation of energy resolving X-ray detectors \cite{van_assche:21} should in combination with suitable ML algorithms be able to mitigate the effect of  beam hardening.

Another approach could be having a neural network learn the packing geometry distribution and then use this knowledge to use less than 10\% of the raw images to create a 3D reconstruction \cite{jin:17}, which paves the way for speeding up data acquisition in scanning techniques. Last but not least, ML algorithm also finds applications in raw data processing (e.g., IQ correction of raw data from Radar tracing). More advanced and efficient ML algorithms will be of great interest in the future development of imaging techniques discussed above.

---------------------------------------------------

\begin{acknowledgements}
We thank Jinchen Zhao, Zhiyun Lu, and Han Yue for helpful discussions, Ra\'ul Cruz Hidalgo for providing the simulation data used in Fig.\,\ref{dynamics}, and Joelle Clau{\ss}en and Stefan Gerth for the permission to use the image in panel a of figure \ref{fig:statics}.  This work is partly supported by the Deutsche Forschungsgemeinschaft through Grant No.~HU1939/4-1 and Startup Grant from Duke Kunshan University. The Collective Dynamics Lab at Duke Kunshan University is partly sponsored by a philanthropic gift.
\end{acknowledgements}

\bibliography{pip}

\begin{thebibliography}{100}

\bibitem{Borges1974}
J.~L. Borges, {\em In praise of darkness}.
\newblock New York: Dutton, 1st ed.~ed., 1974.

\bibitem{Beiser2018}
V.~Beiser, {\em The world in a grain: the story of sand and how it transformed
  civilization}.
\newblock New York: Riverhead Books, 2018.

\bibitem{UEPG2021}
UEPG, ``Annual {Review} of {UEPG} (the {European} {Aggregates}
  {Association}),'' tech. rep., Nov. 2021.

\bibitem{NASA2011}
{NASA}, ``{Avalanche} on {North} pole scarp on {Mars},'' 2011.

\bibitem{Huang2014}
K.~Huang, {\em Wet granular dynamics: From single particle bouncing to
  collective motion}.
\newblock Habilitation thesis, University of Bayreuth, 2014.

\bibitem{Meron2015}
E.~Meron, {\em Nonlinear physics of ecosystems}.
\newblock Boca Raton, FL: CRC Press, Taylor \& Francis Group, 2015.

\bibitem{Antonyuk2009}
S.~Antonyuk, S.~Heinrich, N.~Deen, and H.~Kuipers, ``Influence of liquid layers
  on energy absorption during particle impact,'' {\em Particuology}, vol.~7,
  pp.~245--259, Aug. 2009.

\bibitem{Mueller2016}
T.~M\"uller and K.~Huang, ``Influence of the liquid film thickness on the
  coefficient of restitution for wet particles,'' {\em Physical Review E},
  vol.~93, p.~042904, Apr. 2016.

\bibitem{Amon2017}
A.~Amon, P.~Born, and et~al., ``Preface: {Focus} on imaging methods in granular
  physics,'' {\em Review of Scientific Instruments}, vol.~88, p.~051701, May
  2017.

\bibitem{Rosato2020}
A.~Rosato and K.~Windows-Yule, ``Chapter 3 - {Investigative} approaches {I}:
  experimental imaging techniques,'' in {\em Segregation in {Vibrated}
  {Granular} {Systems}} (A.~Rosato and K.~Windows-Yule, eds.), pp.~37--74,
  Academic Press, Jan. 2020.

\bibitem{Yule2022}
C.~R.~K. Windows-Yule, M.~T. Herald, and et~al., ``Recent advances in positron
  emission particle tracking: a comparative review,'' {\em Reports on Progress
  in Physics}, vol.~85, p.~016101, Jan. 2022.

\bibitem{Andreotti2013}
B.~Andreotti, Y.~Forterre, and O.~Pouliquen, {\em Granular {Media}: {Between}
  {Fluid} and {Solid}}.
\newblock Cambridge University Press, June 2013.

\bibitem{Johnson2005}
P.~A. Johnson and X.~Jia, ``Nonlinear dynamics, granular media and dynamic
  earthquake triggering,'' {\em Nature}, vol.~437, p.~04015, Oct. 2005.

\bibitem{Gray2018}
J.~M. N.~T. Gray, ``Particle {Segregation} in {Dense} {Granular} {Flows},''
  {\em Annual Review of Fluid Mechanics}, vol.~50, no.~1, pp.~407--433, 2018.

\bibitem{Jerolmack2019}
D.~J. Jerolmack and K.~E. Daniels, ``Viewing {Earth}’s surface as a
  soft-matter landscape,'' {\em Nature Reviews Physics}, vol.~1, pp.~716--730,
  Dec. 2019.

\bibitem{ruiz-suarez_penetration_2013}
J.~C. Ruiz-Su\'arez, ``Penetration of projectiles into granular targets,'' {\em
  Reports on Progress in Physics}, vol.~76, no.~6, p.~066601, 2013.

\bibitem{Huang2020}
K.~Huang, D.~Hern\'andez-Delfin, and et~al., ``The role of initial speed in
  projectile impacts into light granular media,'' {\em Scientific Reports},
  vol.~10, pp.~1--12, Feb. 2020.

\bibitem{Feng2022}
Y.~Feng, S.~Huang, and et~al., ``Granular dynamics in auger sampling,'' {\em
  Journal of Fluid Mechanics}, vol.~935, p.~A26, Mar. 2022.

\bibitem{Jaeger1996}
H.~M. Jaeger, S.~R. Nagel, and R.~P. Behringer, ``Granular solids, liquids, and
  gases,'' {\em Reviews of Modern Physics}, vol.~68, p.~1259, Oct. 1996.

\bibitem{Duran2000}
J.~Duran, {\em Sands, {Powders} and {Grains} ({An} {Introduction} to the
  {Physics} of {Granular} {Materials})}.
\newblock New York: Springer-Verlag, 1~ed., 2000.

\bibitem{Haff1983}
P.~K. Haff, ``Grain flow as a fluid-mechanical phenomenon,'' {\em Journal of
  Fluid Mechanics}, vol.~134, pp.~401--430, 1983.

\bibitem{Poeschel2003}
T.~P\"oschel and N.~Brilliantov, {\em Granular gas dynamics}.
\newblock Berlin; New York: Springer, 2003.

\bibitem{Goldhirsch2003}
I.~Goldhirsch, ``Rapid granular flows,'' {\em Annu. Rev. Fluid Mech.}, vol.~35,
  p.~267, 2003.

\bibitem{Eshuis2010}
P.~Eshuis, K.~van~der Weele, D.~Lohse, and D.~van~der Meer, ``Experimental
  {Realization} of a {Rotational} {Ratchet} in a {Granular} {Gas},'' {\em
  Physical Review Letters}, vol.~104, p.~248001, June 2010.

\bibitem{Fullmer2017}
W.~D. Fullmer and C.~M. Hrenya, ``The {Clustering} {Instability} in {Rapid}
  {Granular} and {Gas}-{Solid} {Flows},'' {\em Annual Review of Fluid
  Mechanics}, vol.~49, no.~1, pp.~485--510, 2017.

\bibitem{MiDi2004}
G.~D.~R. MiDi, ``On dense granular flows,'' {\em The European Physical Journal
  E}, vol.~14, no.~4, p.~25, 2004.

\bibitem{Jop2006}
P.~Jop, Y.~Forterre, and O.~Pouliquen, ``A constitutive law for dense granular
  flows,'' {\em Nature}, vol.~441, pp.~727--730, June 2006.

\bibitem{Bak1987}
P.~Bak, C.~Tang, and K.~Wiesenfeld, ``Self-organized criticality: {An}
  explanation of the 1/f noise,'' {\em Physical Review Letters}, vol.~59,
  pp.~381--384, July 1987.

\bibitem{Bak1988}
P.~Bak, C.~Tang, and K.~Wiesenfeld, ``Self-organized criticality,'' {\em
  Physical Review A}, vol.~38, pp.~364--374, July 1988.

\bibitem{Huang2015}
K.~Huang, ``1/f noise on the brink of wet granular melting,'' {\em New Journal
  of Physics}, vol.~17, Aug. 2015.

\bibitem{Edwards1989}
S.~Edwards and R.~Oakeshott, ``Theory of powders,'' {\em Physica A: Statistical
  Mechanics and its Applications}, vol.~157, pp.~1080--1090, June 1989.

\bibitem{Baule2018}
A.~Baule, F.~Morone, H.~J. Herrmann, and H.~Makse, ``Edwards statistical
  mechanics for jammed granular matter,'' {\em Reviews of Modern Physics},
  vol.~90, p.~015006, Mar. 2018.

\bibitem{Henann2013}
D.~L. Henann and K.~Kamrin, ``A predictive, size-dependent continuum model for
  dense granular flows,'' {\em Proceedings of the National Academy of
  Sciences}, vol.~110, pp.~6730--6735, Apr. 2013.

\bibitem{Kamrin2018}
K.~Kamrin, ``Quantitative {Rheological} {Model} for {Granular} {Materials}:
  {The} {Importance} of {Particle} {Size},'' in {\em Handbook of {Materials}
  {Modeling}: {Applications}: {Current} and {Emerging} {Materials}}
  (W.~Andreoni and S.~Yip, eds.), pp.~1--24, Cham: Springer International
  Publishing, 2018.

\bibitem{Clark2020}
A.~H. Clark and J.~A. Dijksman, ``Editorial: {Non}-local {Modeling} and
  {Diverging} {Lengthscales} in {Structured} {Fluids},'' {\em Frontiers in
  Physics}, vol.~8, p.~18, Feb. 2020.

\bibitem{Aguirre2021}
M.~A. Aguirre, S.~Luding, L.~A. Pugnaloni, and R.~Soto, ``Editorial: {Powders}
  \& {Grains} 2021 9th {International} {Conference} on {Micromechanics} of
  {Granular} {Media},'' {\em EPJ Web of Conferences}, vol.~249, p.~00001, 2021.

\bibitem{Poeschel2005}
T.~P\"oschel and T.~Schwager, {\em Computational granular dynamics models and
  algorithms}.
\newblock Berlin: Springer-Verlag, 2005.

\bibitem{Zhao2016}
J.~Zhao, M.~Jiang, K.~Soga, and S.~Luding, ``Micro origins for macro behavior
  in granular media,'' {\em Granular Matter}, vol.~18, p.~59, July 2016.

\bibitem{Ge2017}
W.~Ge, L.~Wang, and et~al., ``Discrete simulation of granular and
  particle-fluid flows: from fundamental study to engineering application,''
  {\em Reviews in Chemical Engineering}, vol.~33, no.~6, pp.~551--623, 2017.

\bibitem{Kieckhefen2020}
P.~Kieckhefen, S.~Pietsch, M.~Dosta, and S.~Heinrich, ``Possibilities and
  {Limits} of {Computational} {Fluid} {Dynamics}-{Discrete} {Element} {Method}
  {Simulations} in {Process} {Engineering}: {A} {Review} of {Recent}
  {Advancements} and {Future} {Trends},'' {\em Annual Review of Chemical and
  Biomolecular Engineering}, vol.~11, no.~1, pp.~397--422, 2020.

\bibitem{Thompson1995}
A.~P. Thompson, H.~M. Aktulga, and et~al., ``{LAMMPS} - a flexible simulation
  tool for particle-based materials modeling at the atomic, meso, and continuum
  scales,'' {\em Computer Physics Communications}, vol.~271, p.~108171, 1995.

\bibitem{Kloss2012}
C.~Kloss, C.~Goniva, A.~Hager, S.~Amberger, and S.~Pirker, ``Models, algorithms
  and validation for opensource {DEM} and {CFD}-{DEM},'' {\em Progress in
  Computational Fluid Dynamics, An International Journal}, vol.~12, no.~2/3,
  p.~140, 2012.

\bibitem{Weinhart2020}
T.~Weinhart, L.~Orefice, and et~al., ``Fast, flexible particle simulations -
  {An} introduction to {MercuryDPM},'' {\em Computer Physics Communications},
  vol.~249, p.~107129, Apr. 2020.

\bibitem{Scheiwiller1987}
T.~Scheiwiller, ``Dynamics of {Powder} {Snow} {Avalanches},'' {\em Annales
  Geophysicae}, vol.~58, p.~569, 1987.

\bibitem{Aranson2017}
I.~S. Aranson and A.~Pikovsky, {\em Advances in {Dynamics}, {Patterns},
  {Cognition} : {Challenges} in {Complexity}}.
\newblock Nonlinear {Systems} and {Complexity}, 2195-9994 ; 20, Cham : Springer
  International Publishing : Imprint: Springer, 2017., 2017.

\bibitem{Agudo:17}
J.~R. Agudo, G.~Luzi, J.~Han, M.~Hwang, J.~Lee, and A.~Wierschem, ``Detection
  of particle motion using image processing with particular emphasis on rolling
  motion,'' {\em Review of Scientific Instruments}, vol.~88, p.~051805, 2017.

\bibitem{Amon2017_DWS}
A.~Amon, A.~Mikhailovskaya, and J.~Crassous, ``Spatially resolved measurements
  of micro-deformations in granular materials using diffusing wave
  spectroscopy,'' {\em Review of Scientific Instruments}, vol.~88, p.~051804,
  May 2017.

\bibitem{Born2017}
P.~Born and K.~Holldack, ``Analysis of granular packing structure by scattering
  of {THz} radiation,'' {\em Review of Scientific Instruments}, vol.~88,
  p.~051802, May 2017.

\bibitem{daniels:17}
K.~E. Daniels, J.~E. Kollmer, and J.~G. Puckett, ``Photoelastic force
  measurements in granular materials,'' {\em Review of Scientific Instruments},
  vol.~88, p.~051808, 2017.

\bibitem{Dijksman2017}
J.~A. Dijksman, N.~Brodu, and R.~P. Behringer, ``Refractive index matched
  scanning and detection of soft particles,'' {\em Review of Scientific
  Instruments}, vol.~88, p.~051807, May 2017.

\bibitem{Ott2017}
F.~Ott, S.~Herminghaus, and K.~Huang, ``Radar for tracer particles,'' {\em
  Review of Scientific Instruments}, vol.~88, p.~051801, May 2017.

\bibitem{Parker2017}
D.~J. Parker, ``Positron emission particle tracking and its application to
  granular media,'' {\em Review of Scientific Instruments}, vol.~88, p.~051803,
  May 2017.

\bibitem{Stannarius2017}
R.~Stannarius, ``Magnetic resonance imaging of granular materials,'' {\em
  Review of Scientific Instruments}, vol.~88, p.~051806, May 2017.

\bibitem{Weis2017}
S.~Weis and M.~Schr\"oter, ``Analyzing {X}-ray tomographies of granular
  packings,'' {\em Review of Scientific Instruments}, vol.~88, no.~5,
  p.~051809, 2017.

\bibitem{Han2016}
E.~Han, I.~R. Peters, and H.~M. Jaeger, ``High-speed ultrasound imaging in
  dense suspensions reveals impact-activated solidification due to dynamic
  shear jamming,'' {\em Nature Communications}, vol.~7, p.~12243, July 2016.

\bibitem{brzinski:18}
T.~A. Brzinski and K.~E. Daniels, ``Sounds of {Failure}: {Passive} {Acoustic}
  {Measurements} of {Excited} {Vibrational} {Modes},'' {\em Physical Review
  Letters}, vol.~120, p.~218003, 2018.

\bibitem{bassett:12}
D.~S. Bassett, E.~T. Owens, K.~E. Daniels, and M.~A. Porter, ``Influence of
  network topology on sound propagation in granular materials,'' {\em Physical
  Review E}, vol.~86, p.~041306, 2012.

\bibitem{Windows-Yule2022}
C.~R.~K. Windows-Yule, M.~T. Herald, and et~al., ``Recent advances in positron
  emission particle tracking: a comparative review,'' {\em Reports on Progress
  in Physics}, vol.~85, p.~016101, Jan. 2022.

\bibitem{Lin1985}
J.~S. Lin, M.~M. Chen, and B.~T. Chao, ``A novel radioactive particle tracking
  facility for measurement of solids motion in gas fluidized beds,'' {\em AIChE
  Journal}, vol.~31, pp.~465--473, Mar. 1985.

\bibitem{Neuwirth2013}
J.~Neuwirth, S.~Antonyuk, S.~Heinrich, and M.~Jacob, ``{CFD}–{DEM} study and
  direct measurement of the granular flow in a rotor granulator,'' {\em
  Chemical Engineering Science}, vol.~86, pp.~151--163, Feb. 2013.

\bibitem{tao:20}
X.~Tao and H.~Wu, ``The translational and rotational motions of a cylindrical
  particle in a granular shear flow inside a split bottom {Couette} cell,''
  {\em Physics of Fluids}, vol.~32, p.~073310, 2020.

\bibitem{Rech2020}
F.~Rech and K.~Huang, ``Radar for projectile impact on granular media,'' {\em
  International Journal of Microwave and Wireless Technologies}, vol.~12,
  pp.~1--7, May 2020.

\bibitem{gerth:21}
S.~Gerth, J.~Clau{\ss}en, and et~al., ``Semiautomated {3D} {Root}
  {Segmentation} and {Evaluation} {Based} on {X}-{Ray} {CT} {Imagery},'' {\em
  Plant Phenomics}, vol.~2021, 2021.

\bibitem{baur:19}
M.~Baur, N.~Uhlmann, T.~P\"oschel, and M.~Schr\"oter, ``Correction of beam
  hardening in {X}-ray radiograms,'' {\em Review of Scientific Instruments},
  vol.~90, p.~025108, 2019.

\bibitem{ando_13}
E.~And\`o, G.~Viggiani, S.~A. Hall, and J.~Desrues, ``Experimental
  micro-mechanics of granular media studied by x-ray tomography: recent results
  and challenges,'' {\em G\'eotechnique Letters}, vol.~3, pp.~142--146, 2013.

\bibitem{okubadejo_17}
{\em Identification and tracking of particles undergoing progressive breakage
  under stress with {3D}+ t image analysis}, 3rd International Conference on
  Tomography of Materials and Structures, Lund, Sweden, 2017.

\bibitem{ando_19}
E.~And\`o, J.~Dijkstra, E.~Roubin, C.~Dano, and E.~Boller, ``A peek into the
  origin of creep in sand,'' {\em Granular Matter}, vol.~21, p.~11, 2019.

\bibitem{viggiani_14}
G.~Viggiani, E.~And\`o, D.~Takano, and J.~C. Santamarina, ``Laboratory {X}-ray
  {Tomography}: {A} {Valuable} {Experimental} {Tool} for {Revealing}
  {Processes} in {Soils},'' {\em Geotechnical Testing Journal}, vol.~38,
  p.~20140060, 2014.

\bibitem{martins_20}
A.~D. Martins, F.~O'Callaghan, A.~G. Bengough, K.~W. Loades, M.~Pasqual,
  E.~Kolb, and L.~X. Dupuy, ``The helical motions of roots are linked to
  avoidance of particle forces in soil,'' {\em New Phytologist}, vol.~225,
  pp.~2356--2367, 2020.

\bibitem{buzug:08}
T.~M. Buzug, {\em Computed Tomography}.
\newblock Berlin: Springer, 2008.

\bibitem{hemmerle:16}
A.~Hemmerle, M.~Schr\"oter, and L.~Goehring, ``A cohesive granular material
  with tunable elasticity,'' {\em Scientific Reports}, vol.~6, p.~35650, 2016.

\bibitem{li:21}
Z.~Li, Z.~Zeng, Y.~Xing, J.~Li, J.~Zheng, Q.~Mao, J.~Zhang, M.~Hou, and
  Y.~Wang, ``Microscopic structure and dynamics study of granular segregation
  mechanism by cyclic shear,'' {\em Science Advances}, vol.~7, p.~eabe8737,
  2021.

\bibitem{yuan:21}
Y.~Yuan, Y.~Xing, J.~Zheng, Z.~Li, H.~Yuan, S.~Zhang, Z.~Zeng, C.~Xia, H.~Tong,
  W.~Kob, J.~Zhang, and Y.~Wang, ``Experimental {Test} of the {Edwards}
  {Volume} {Ensemble} for {Tapped} {Granular} {Packings},'' {\em Physical
  Review Letters}, vol.~127, p.~018002, 2021.

\bibitem{boerzsoenyi:12}
T.~B\"orzs\"onyi, B.~Szab\'o, and et~al., ``Shear-induced alignment and
  dynamics of elongated granular particles,'' {\em Physical Review E}, vol.~86,
  p.~051304, 2012.

\bibitem{xia:14}
C.~Xia, K.~Zhu, Y.~Cao, H.~Sun, B.~Kou, and Y.~Wang, ``X-ray tomography study
  of the random packing structure of ellipsoids,'' {\em Soft Matter}, vol.~10,
  pp.~990--996, 2014.

\bibitem{wang:14}
Y.~Wang, C.~Xia, Y.~Cao, B.~Kou, J.~Li, X.~Xiao, and K.~Fezzaa, ``Fast x-ray
  micro-tomography imaging study of granular packing under tapping,'' in {\em
  Developments in {X}-{Ray} {Tomography} {IX}}, vol.~9212, pp.~121--131, SPIE,
  2014.

\bibitem{waktola:18}
S.~Waktola, A.~Bieberle, F.~Barthel, M.~Bieberle, U.~Hampel, K.~Grudzie\'n, and
  L.~Babout, ``A new data-processing approach to study particle motion using
  ultrafast {X}-ray tomography scanner: case study of gravitational mass
  flow,'' {\em Experiments in Fluids}, vol.~59, p.~69, 2018.

\bibitem{stannarius:19}
R.~Stannarius, D.~S. Martinez, T.~B\"orzs\"onyi, M.~Bieberle, F.~Barthel, and
  U.~Hampel, ``High-speed x-ray tomography of silo discharge,'' {\em New
  Journal of Physics}, vol.~21, p.~113054, 2019.

\bibitem{homan:15}
T.~Homan, R.~Mudde, D.~Lohse, and D.~v.~d. Meer, ``High-speed {X}-ray imaging
  of a ball impacting on loose sand,'' {\em Journal of Fluid Mechanics},
  vol.~777, pp.~690--706, 2015.

\bibitem{khalili:17}
M.~H. Khalili, S.~Brisard, M.~Bornert, P.~Aimedieu, J.-M. Pereira, and J.-N.
  Roux, ``Discrete {Digital} {Projections} {Correlation}: {A}
  {Reconstruction}-{Free} {Method} to {Quantify} {Local} {Kinematics} in
  {Granular} {Media} by {X}-ray {Tomography},'' {\em Experimental Mechanics},
  vol.~57, pp.~819--830, 2017.

\bibitem{gupta:21}
A.~Gupta, R.~S. Crum, C.~Zhai, K.~T. Ramesh, and R.~C. Hurley, ``Quantifying
  particle-scale {3D} granular dynamics during rapid compaction from
  time-resolved in situ {2D} x-ray images,'' {\em Journal of Applied Physics},
  vol.~129, p.~225902, 2021.

\bibitem{ando:21}
E.~And\`o, B.~Marks, and S.~Roux, ``Single-projection reconstruction technique
  for positioning monodisperse spheres in {3D} with a divergent x-ray beam,''
  {\em Measurement Science and Technology}, vol.~32, p.~095405, 2021.

\bibitem{berger_nist_2010}
M.~Berger, J.~Hubbell, and et~al., ``{NIST} {XCOM}: {Photon} {Cross} {Sections}
  {Database} - {Version} {History},'' 2010.

\bibitem{kollmer:20}
J.~E. Kollmer, T.~Shreve, and et~al., ``Migrating {Shear} {Bands} in {Shaken}
  {Granular} {Matter},'' {\em Physical Review Letters}, vol.~125, p.~048001,
  2020.

\bibitem{xia:15}
C.~Xia, J.~Li, Y.~Cao, B.~Kou, X.~Xiao, K.~Fezzaa, T.~Xiao, and Y.~Wang, ``The
  structural origin of the hard-sphere glass transition in granular packing,''
  {\em Nature Communications}, vol.~6, p.~8409, 2015.

\bibitem{neudecker:13}
M.~Neudecker, S.~Ulrich, S.~Herminghaus, and M.~Schr\"oter, ``Jammed
  {Frictional} {Tetrahedra} are {Hyperstatic},'' {\em Physical Review Letters},
  vol.~111, p.~028001, 2013.

\bibitem{xu:21}
Z.~Xu, J.~Yang, Y.~Ding, Y.~Zhao, J.~Li, B.~Hu, and C.~Xia, ``Packing and void
  structures of octahedral, dodecahedral and icosahedral granular particles,''
  {\em Granular Matter}, vol.~23, p.~88, 2021.

\bibitem{vlahinic:14}
I.~Vlahini\'c, E.~And\'o, G.~Viggiani, and J.~E. Andrade, ``Towards a more
  accurate characterization of granular media: extracting quantitative
  descriptors from tomographic images,'' {\em Granular Matter}, vol.~16,
  pp.~9--21, 2014.

\bibitem{wiebicke:17}
M.~Wiebicke, E.~And\'o, I.~Herle, and G.~Viggiani, ``On the metrology of
  interparticle contacts in sand from x-ray tomography images,'' {\em
  Measurement Science and Technology}, vol.~28, p.~124007, 2017.

\bibitem{bull:20}
D.~J. Bull, J.~A. Smethurst, I.~Sinclair, F.~Pierron, T.~Roose, W.~Powrie, and
  A.~G. Bengough, ``Mechanisms of root reinforcement in soils: an experimental
  methodology using four-dimensional {X}-ray computed tomography and digital
  volume correlation,'' {\em Proceedings of the Royal Society A: Mathematical,
  Physical and Engineering Sciences}, vol.~476, p.~20190838, 2020.

\bibitem{anselmucci_21}
F.~Anselmucci, E.~And\`o, G.~Viggiani, N.~Lenoir, C.~Arson, and L.~Sibille,
  ``Imaging local soil kinematics during the first days of maize root growth in
  sand,'' {\em Scientific Reports}, vol.~11, p.~22262, 2021.

\bibitem{weis:19}
S.~Weis, G.~E. Schr\"oder-Turk, and M.~Schr\"oter, ``Structural similarity
  between dry and wet sphere packings,'' {\em New Journal of Physics}, vol.~21,
  p.~043020, 2019.

\bibitem{hiller:19}
T.~Hiller, J.~Ardevol-Murison, A.~Muggeridge, M.~Schr\"oter, and M.~Brinkmann,
  ``The {Impact} of {Wetting}-{Heterogeneity} {Distribution} on {Capillary}
  {Pressure} and {Macroscopic} {Measures} of {Wettability},'' {\em SPE
  Journal}, vol.~24, pp.~200--214, 2019.

\bibitem{Brodu2015}
N.~Brodu, J.~A. Dijksman, and R.~P. Behringer, ``Spanning the scales of
  granular materials through microscopic force imaging,'' {\em Nature
  Communications}, vol.~6, p.~6361, May 2015.

\bibitem{hurley:16}
R.~Hurley, S.~Hall, J.~Andrade, and J.~Wright, ``Quantifying {Interparticle}
  {Forces} and {Heterogeneity} in {3D} {Granular} {Materials},'' {\em Physical
  Review Letters}, vol.~117, p.~098005, 2016.

\bibitem{Hurley2022}
R.~Hurley and C.~Zhai, ``Challenges and opportunities in measuring
  time-resolved force chain evolution in {3D} granular materials,'' {\em Papers
  in Physics}, vol.~14, p.~140003, Mar. 2022.

\bibitem{Skolnik2008}
M.~I. Skolnik, ed., {\em Radar handbook}.
\newblock New York: McGraw-Hill, 3rd ed~ed., 2008.

\bibitem{Saponara2019}
S.~Saponara, M.~S. Greco, and F.~Gini, ``Radar-on-{Chip}/in-{Package} in
  {Autonomous} {Driving} {Vehicles} and {Intelligent} {Transport} {Systems}:
  {Opportunities} and {Challenges},'' {\em IEEE Signal Processing Magazine},
  vol.~36, pp.~71--84, Sept. 2019.
\newblock Conference Name: IEEE Signal Processing Magazine.

\bibitem{schroeter:06}
M.~Schr\"oter, S.~Ulrich, J.~Kreft, J.~B. Swift, and H.~L. Swinney,
  ``Mechanisms in the size segregation of a binary granular mixture,'' {\em
  Physical Review E}, vol.~74, p.~011307, 2006.

\bibitem{Umbanhowar2019}
P.~B. Umbanhowar, R.~M. Lueptow, and J.~M. Ottino, ``Modeling {Segregation} in
  {Granular} {Flows},'' {\em Annual Review of Chemical and Biomolecular
  Engineering}, vol.~10, no.~1, pp.~129--153, 2019.

\bibitem{Rasmussen2006}
C.~E. Rasmussen and C.~K.~I. Williams, {\em Gaussian processes for machine
  learning}.
\newblock Adaptive computation and machine learning, Cambridge, Mass: MIT
  Press, 2006.
\newblock OCLC: ocm61285753.

\bibitem{Altshuler2014}
E.~Altshuler, H.~Torres, A.~González-Pita, G.~Sánchez-Colina,
  C.~Pérez-Penichet, S.~Waitukaitis, and R.~C. Hidalgo, ``Settling into dry
  granular media in different gravities,'' {\em Geophysical Research Letters},
  vol.~41, pp.~3032--3037, May 2014.

\bibitem{Koestler2021}
S.~Köstler, J.~Zhao, C.~Lyu, S.~Völkel, and K.~Huang, ``Embedded inertial
  sensor for tracking projectile impact on granular media,'' {\em EPJ Web of
  Conferences}, vol.~249, p.~15007, 2021.

\bibitem{Swarm2019}
{CORDIS}, ``How to explore inaccessible places by swarms of sensors,'' 2019.

\bibitem{bizon:98}
C.~Bizon, M.~D. Shattuck, J.~B. Swift, W.~D. McCormick, and H.~L. Swinney,
  ``Patterns in {3D} {Vertically} {Oscillated} {Granular} {Layers}:
  {Simulation} and {Experiment},'' {\em Physical Review Letters}, vol.~80,
  pp.~57--60, 1998.

\bibitem{boerzsoenyi:09}
T.~B\"orzs\"onyi, R.~E. Ecke, and J.~N. McElwaine, ``Patterns in {Flowing}
  {Sand}: {Understanding} the {Physics} of {Granular} {Flow},'' {\em Physical
  Review Letters}, vol.~103, p.~178302, 2009.

\bibitem{butzhammer2015}
L.~Butzhammer, S.~V\"olkel, I.~Rehberg, and K.~Huang, ``Pattern formation in
  wet granular matter under vertical vibrations,'' {\em Physical Review E},
  vol.~92, p.~012202, July 2015.

\bibitem{tsukahara:08}
M.~Tsukahara, S.~Mitrovic, V.~Gajdosik, G.~Margaritondo, L.~Pournin,
  M.~Ramaioli, D.~Sage, Y.~Hwu, M.~Unser, and T.~M. Liebling, ``Coupled
  tomography and distinct-element-method approach to exploring the granular
  media microstructure in a jamming hourglass,'' {\em Physical Review E},
  vol.~77, p.~061306, 2008.

\bibitem{bannerman:11}
M.~N. Bannerman, J.~E. Kollmer, A.~Sack, M.~Heckel, P.~Mueller, and
  T.~P\"oschel, ``Movers and shakers: {Granular} damping in microgravity,''
  {\em Physical Review E}, vol.~84, p.~011301, 2011.

\bibitem{Huang2018}
K.~Huang, ``A {Hard}-{Sphere} {Model} for {Wet} {Granular} {Dynamics},'' in
  {\em Proceedings of {China}-{Europe} {Conference} on {Geotechnical}
  {Engineering}} (W.~Wu and H.-S. Yu, eds.), Springer {Series} in
  {Geomechanics} and {Geoengineering}, (Cham), pp.~169--173, Springer
  International Publishing, 2018.

\bibitem{Wu2018}
W.~Wu and H.-S. Yu, eds., {\em Proceedings of {China}-{Europe} {Conference} on
  {Geotechnical} {Engineering}: {Volume} 2}.
\newblock Springer {Series} in {Geomechanics} and {Geoengineering}, Cham:
  Springer International Publishing : Imprint: Springer, 1st ed. 2018~ed.,
  2018.

\bibitem{moreno-atanasio:10}
R.~Moreno-Atanasio, R.~A. Williams, and X.~Jia, ``Combining {X}-ray
  microtomography with computer simulation for analysis of granular and porous
  materials,'' {\em Particuology}, vol.~8, pp.~81--99, 2010.

\bibitem{Dosta2018}
M.~Dosta, U.~Br\"ockel, L.~Gilson, S.~Kozhar, G.~K. Auernhammer, and
  S.~Heinrich, ``Application of micro computed tomography for adjustment of
  model parameters for discrete element method,'' {\em Chemical Engineering
  Research and Design}, vol.~135, pp.~121--128, July 2018.

\bibitem{Puzyrev2020}
D.~Puzyrev, K.~Harth, T.~Trittel, and R.~Stannarius, ``Machine {Learning} for
  {3D} {Particle} {Tracking} in {Granular} {Gases},'' {\em Microgravity Science
  and Technology}, vol.~32, pp.~897--906, Oct. 2020.

\bibitem{nicusan:20}
A.~L. Nicu\c{c}an and C.~R.~K. Windows-Yule, ``Positron emission particle
  tracking using machine learning,'' {\em Review of Scientific Instruments},
  vol.~91, p.~013329, 2020.

\bibitem{lehtinen:18}
J.~Lehtinen, J.~Munkberg, J.~Hasselgren, S.~Laine, T.~Karras, M.~Aittala, and
  T.~Aila, ``{Noise2Noise}: {Learning} {Image} {Restoration} without {Clean}
  {Data},'' {\em arXiv:1803.04189 [cs, stat]}, 2018.
\newblock arXiv: 1803.04189.

\bibitem{cong:21}
W.~Cong, Y.~Xi, B.~D. Man, and G.~Wang, ``Monochromatic image reconstruction
  via machine learning,'' {\em Machine Learning: Science and Technology},
  vol.~2, p.~025032, 2021.

\bibitem{van_assche:21}
F.~Van~Assche, S.~Vanheule, L.~Van~Hoorebeke, and M.~N. Boone, ``The {Spectral}
  {X}-ray {Imaging} {Data} {Acquisition} ({SpeXIDAQ}) {Framework},'' {\em
  Sensors}, vol.~21, p.~563, 2021.

\bibitem{jin:17}
K.~H. Jin, M.~T. McCann, E.~Froustey, and M.~Unser, ``Deep {Convolutional}
  {Neural} {Network} for {Inverse} {Problems} in {Imaging},'' {\em IEEE
  Transactions on Image Processing}, vol.~26, pp.~4509--4522, 2017.

\end{thebibliography}
\bibliographystyle{ieeetr}

\end{document}